\documentclass[10pt]{article}
\usepackage{amsmath}
\usepackage{graphicx}
\usepackage{color}
\usepackage{orcidlink}
\usepackage{hyperref}
\hypersetup{colorlinks=true, linkcolor=blue, citecolor=blue, urlcolor=blue}
\usepackage{subcaption}
\begin{document}
\baselineskip=14 pt

\begin{center}
{\large{\bf Study of scalar particles through the Klein-Gordon equation under rainbow gravity effects in Bonnor-Melvin-Lambda space-time}}
\end{center}

\vspace{0.3cm}

\begin{center}
    {\bf Faizuddin Ahmed\orcidlink{0000-0003-2196-9622}}\footnote{\bf faizuddinahmed15@gmail.com ; faizuddin@ustm.ac.in}\\
    \vspace{0.1cm}
    {\it Department of Physics, University of Science \& Technology Meghalaya, Ri-Bhoi, 793101, India}\\
    \vspace{0.3cm}
    {\bf Abdelmalek Bouzenada\orcidlink{0000-0002-3363-980X}}\footnote{\bf abdelmalek.bouzenada@univ-tebessa.dz ; abdelmalekbouzenada@gmail.com}\\
    \vspace{0.1cm}
    {\it  Laboratory of theoretical and applied Physics, Echahid Cheikh Larbi Tebessi University, Algeria}\\
\end{center}

\vspace{0.3cm}

\begin{abstract}
In our investigation, we explore the quantum dynamics of charge-free scalar particles through the Klein-Gordon equation within the framework of rainbow gravity's, considering the Bonnor-Melvin-Lambda (BML) space-time background. The BML solution is characterized by the magnetic field strength along the axis of symmetry direction which is related with the cosmological constant $\Lambda$ and the topological parameter $\alpha$ of the geometry. The behavior of charge-free scalar particles described by the Klein-Gordon equation is investigated, utilizing two sets of rainbow functions: (i) $f(\chi)=\frac{(e^{\beta\,\chi}-1)}{\beta\,\chi}$,\, $h(\chi)=1$ and (ii) $f(\chi)=1$,\, $h(\chi)=1+\frac{\beta\,\chi}{2}$. Here $0 < \Big(\chi=\frac{|E|}{E_p}\Big) \leq 1$ with $E$ represents the particle's energy, $E_p$ is the Planck's energy, and $\beta$ is the rainbow parameter. We obtain the approximate analytical solutions for the scalar particles and conduct a thorough analysis of the obtained results. Afterwards, we study the quantum dynamics of quantum oscillator fields within this BML space-time, employing the Klein-Gordon oscillator. Here also, we choose the same sets of rainbow functions and obtained approximate eigenvalue solution for the oscillator fields. Notably, we demonstrate that the relativistic approximate energy profiles of charge-free scalar particles and oscillator fields get influenced by the topology of the geometry and the cosmological constant. Furthermore, we show that the energy profiles of scalar particles get modifications by the rainbow parameter and the quantum oscillator fields by both the rainbow parameter and the frequency of oscillation.
\end{abstract}

\vspace{0.1cm}


\vspace{0.1cm}


\section{Introduction}
\label{intro}

Starting on a captivating journey into the intricate dance between gravitational forces and the dynamics of quantum mechanical systems opens up a world of profound exploration. Albert Einstein's revolutionary general theory of relativity (GR) brilliantly envisions gravity as an inherent geometric aspect of space-time \cite{k1}. This groundbreaking theory not only connects space-time curvature with the formation of classical gravitational fields but also yields precise predictions for mesmerizing phenomena such as gravitational waves \cite{k2} and black holes \cite{k3}.

Simultaneously, the robust framework of quantum mechanics (QM) \cite{k4} provides invaluable insights into the nuanced behaviors of particles at the microscopic scale. As these two foundational theories converge, an invitation is extended to delve into the profound mysteries nestled at the crossroads of the macroscopic domain governed by gravity and the quantum intricacies of the subatomic realm. This intersection offers a rich tapestry of scientific inquiry, promising to unravel the secrets that bind the vast cosmos with the smallest building blocks of nature.

In the absence of a definitive theory of quantum gravity (QG), physicists resort to employing semi-classical approaches to tackle the challenges posed by this elusive realm. While these approaches fall short of providing a comprehensive solution, they offer valuable insights into phenomena associated with exceedingly high-energy physics and the early universe \cite{k5,k6,k7,k8,k9,k10,k11,k12}. An illustrative example of such a phenomenological or semi-classical approach involves the violation of Lorentz invariance, wherein the ordinary relativistic dispersion relation is altered by modifying the physical energy and momentum at the Planck scale \cite{k13}. This departure from the dispersion relation has found applications in diverse domains, such as space-time foam models \cite{k14}, loop quantum gravity (QG) \cite{k15}, spontaneous symmetry breaking of Lorentz invariance in string field theory \cite{k16}, spin networks \cite{k17}, discrete space-time \cite{k18}, as well as non-commutative geometry and Lorentz invariance violation \cite{k19}. Subsequently, scientists have extensively explored the myriad applications of rainbow gravity across various physics domains, spanning topics including the isotropic quantum cosmological perfect fluid model within the framework of rainbow gravity \cite{k20}, the adaptation of the Friedmann–Robertson–Walker universe in the context of Einstein-massive rainbow gravity \cite{k21}, the thermodynamics governing black holes \cite{k22}, the geodesic structure characterizing the Schwarzschild black hole \cite{k23,k24}, and the nuanced examination of the massive scalar field in the presence of the Casimir effect \cite{k25}.

The exploration of a coherent framework to comprehend and elucidate phenomena involving high-energy gravitational interactions has captivated the attention of theoretical physicists over the past few decades. An illustrative example of such pursuit is Rainbow gravity, a semi-classical approach that posits the local breakdown of Lorentz symmetry at energy scales akin to the Planck scale $ \left(E_{p}\right) $. Rainbow gravity can be viewed as an extension of the concept of Doubly Special Relativity \cite{k8,k9,k14,k26,k27}. A fundamental facet of this framework is the modification of the metric, contingent upon the ratio of a test particle's energy to the Planck energy, resulting in significant corrections to the energy-momentum dispersion relation. This modification of the relativistic dispersion relation finds motivation in the observation of high-energy cosmic rays \cite{k8}, TeV photons emitted during Gamma Ray Bursts \cite{k14,k28,k29}, and neutrino data from Ice-Cube \cite{k30}. The rainbow gravity effects have been studied in various physical aspects reported in Refs. \cite{BH1, BH2, BH3, BH4, BH5, BH6, BH7, BH8, BH9, BH10, BH11}. 

Following Einstein's proposal of general relativity in 1915, attempts had been made to construct exact solutions to the field equations. The pioneering solution was the renowned Schwarzschild black hole solution. Subsequent advancements included the introduction of de Sitter space and anti-de Sitter space. In 1949, the G\"{o}del cosmological rotating universe was presented, notable for its distinctive characteristic of closed causal curves. Addressing the Einstein-Maxwell equations, Bonnor formulated an exact static solution, discussed in detail for its physical implications \cite{Mel1}. Melvin later revisited this solution, leading to the currently recognized Bonnor-Melvin magnetic universe \cite{Mel2}. An axisymmetric Einstein-Maxwell solution, incorporating a varying magnetic field and a cosmological constant, was constructed in \cite{Mel3}. This electrovacuum solution was subsequently expanded upon in \cite{Mel3,Mel4}. This analysis primary focus on Bonnor-Melvin-type universe featuring a cosmological constant, discussed in detailed in Ref. \cite{MZ}. The specific line-element governing this BML universe with a cosmic string is given by \cite{MZ} ($\hbar=c=G=1$) 
\begin{equation}
{\sf ds^{2}=-dt^{2}+dz^2}+\frac{1}{2\,\Lambda}\,\Big({\sf dr}^{2}+\alpha^{2}\,\sin^{2} r\,{\sf d\phi^{2}}\Big),\label{a1}
\end{equation}
where $\Lambda$ denotes the cosmological constant, and $\alpha$ represents the topological defect parameter which produces an angular deficit by an amount $\Delta\phi=2\,\pi\,(1-\alpha)$. Noted that the cosmic string is introduced into the above line-element by modifying the angular coordinates $\phi \to \alpha\,\phi$, where $0 < \alpha <1$. The strength of the magnetic field for the above space-time geometry  is given by $\mathcal{H} (r)=\frac{\alpha}{\sqrt{2}}\,\sin r $.

Now, we introduce rainbow functions $f(\chi), h(\chi)$ into the Bonnor-Melvin magnetic solution (\ref{a1}) by replacing ${\sf dt} \to \frac{{\sf dt}}{f(\chi)}$ and ${\sf dx}^{i} \to \frac{{\sf dx}^{i}}{h(\chi)}$. Here, $\chi=\frac{|E|}{E_p}$ with $E$ is the particle's energy, and $E_p$ is the Planck's energy and lies in the range $0 < \chi \leq 1$. Therefore, modified line-element of BML space-time (\ref{a1}) under rainbow gravity's is described by the following space-time  
\begin{equation}
{\sf ds^{2}}=-\frac{{\sf dt}^{2}}{f^2 (\chi)}+\frac{1}{h^2 (\chi)}\Bigg[{\sf dz}^2+\frac{1}{2\,\Lambda}\,\Big({\sf dr}^2+\alpha^{2}\,\sin^{2} r\,{\sf d\phi}^{2}\Big)\Bigg].\label{a2}
\end{equation}
One can evaluate the magnetic field strength for the modified BML space-time (\ref{a2}) and it is given by $\mathcal{H} (r)=\frac{\alpha}{\sqrt{2}\,h(\chi)}\,\sin r $ which vanishes on the symmetry axis $r=0$. In the limit $f \to 1$ and $h \to 1$, we will get back the original BML magnetic space-time with a cosmic string given in Eq. (\ref{a1}).

The exploration of relativistic quantum dynamics in a curved space-time background has yielded profound insights into the behavior of various particles, including spin-0 charge-free and charged scalar particles, spin-1/2 fermionic fields, and relativistic spin-1 fields. The outcomes of these studies stand in stark contrast to those obtained in the flat space, showcasing the significant impact of curved space environment. Numerous researchers have also introduced external magnetic and scalar potentials, such as linear confining, Coulomb-type, Cornell-type, Yukawa potential etc., into the quantum systems. These additions have led to intriguing findings, expanding our understanding of how different potentials influence the behavior of quantum particle in curved space-time. Moreover, the presence of topological defects induced by cosmic strings, global monopoles, and spinning cosmic strings has been considered. These defects introduce shifts in the energy spectrum of quantum particles in the quantum realm, adding a layer of complexity to the study of quantum systems in the background of curved space-time and topological structures. Numerous authors have been studied spin-0 scalar particles, spin-1/2 particles in curved space-time background, such as G\"{o}del and G\"{o}del-type solutions, topologically trivial and non-trivial space-times. In addition, investigated have been carried out in the context of topological defect, such as cosmic string space-time, point-like global monopole, cosmic string space-time with spacelike dislocation, screw dislocation etc.. For examples, investigation of scalar charged particlesthrough the Klein-Gordon equation and the fermionic fields in cosmic string space-time in the presence of magnetic field and scalar potential \cite{aa4}, scalar particles through the Duffin-Kemmer-Petiau (DKP) equation in cosmic string background \cite{aa5}, rotating frame effects on scalar fields  through the Klein-Gordon equation in topological defect space-time \cite{aa11} and in cosmic string space-time \cite{bb1}, and the Dirac oscillator in cosmic string space-time in the context of gravity’s rainbow in \cite{aa13}. Furthermore, the relativistic quantum dynamics of scalar particles through the DKP equation have been studied in different curved space-times background without and with topological defect as reported in Refs. \cite{CC1, CC2, CC3, CC4, CC5, CC6, CC7, CC8, CC9, CC10, rev0}. In addition, The generalized Dirac oscillator under the influence of Aharonov-Casher effect in cosmic string space-time \cite{rev1} and under the Aharonov-Bohm effect in a cosmic dislocation space-time \cite{rev2} have also been investigated. 

Our motivation is to study quantum motion of charge-free scalar particles described by the Klein-Gordon equation within the context of rainbow gravity's in the background BML space-time (\ref{a2}) which hasn't yet been studied in quantum systems. Afterwards, we study quantum oscillator fields via the Klein-Gordon oscillator in the same geometry background taking into the rainbow gravity effects. In both scenario, we derive the radial equation of the relativistic wave equation using a suitable wave function ansatz and achieved a homogeneous second-order differential equation. We employ approximation scheme appeared in the radial equation and solve it through special functions. In this analysis, we choose two sets of rainbow function given by: (i) $f(\chi)=\frac{({\sf e}^{\beta\,\chi}-1)}{\beta\,\chi}$,\, $h(\chi)=1$ \cite{k14} and (ii) $f(\chi)=1$,\, $h(\chi)=\Big(1+\frac{\beta\,\chi}{2}\Big)$ \cite{AFG} with $\beta$ is the rainbow parameter. In fact, we show that the energy profiles obtain in both investigations are influenced by the cosmological constant and the topology of the geometry which produces an angular deficit analogue to the cosmic string. This paper is designed as follows: In {\it section 2}, we study quantum dynamics of scalar particles in the background of modified BML space-time under the rainbow gravity's. In {\it section 3}, we study quantum oscillator fields in the background of same space-time and obtain the approximate eigenvalue solutions in both section. In {\it section 4}, we present our conclusions.

\section{Quantum Motions of Scalar particles: The Klein-Gordon Equation}

In this section, we study the quantum motions of charge-free scalar particles under the influence of rainbow gravity's in BML space-time background. We derive the radial equation and solve it through special functions. Therefore, the relativistic quantum dynamics of scalar particles is described by the following relativistic wave equation \cite{aa4,aa5,aa11,bb1,WG}
\begin{eqnarray}
    \Bigg[-\frac{1}{\sqrt{-g}}\,\partial_{\mu}\,\Big(\sqrt{-g}\,g^{\mu\nu}\,\partial_{\nu}\Big)+M^2\Bigg]\,\Psi=0,\label{b1}
\end{eqnarray}
where $M$ is the rest mass of the particles, $g$ is the determinant of the metric tensor $g_{\mu\nu}$ with its inverse $g^{\mu\nu}$. 

The covariant ($g_{\mu\nu}$) and contravariant form ($g_{\mu\nu}$) of the metric tensor for the space-time (\ref{a4}) are given by
\begin{eqnarray}
g_{\mu\nu}=\begin{pmatrix}
-\frac{1}{f^2 (\chi)} & 0 & 0 & 0\\
0 & \frac{1}{2\,\Lambda\,h^2 (\chi)} & 0 & 0\\
0 & 0 & \frac{\alpha^2\,\sin^2 r}{2\,\Lambda\,h^2 (\chi)} & 0\\
0 & 0 & 0 & \frac{1}{h^2 (\chi)}
\end{pmatrix},\quad 
g^{\mu\nu}=\begin{pmatrix}
-f^2(\chi) & 0 & 0 & 0\\
0 & 2\,\Lambda\,h^2(\chi) & 0 & 0\\
0 & 0 & \frac{2\,\Lambda\,h^2 (\chi)}{\alpha^2\,\sin^2 r} & 0\\
0 & 0 & 0 & h^2 (\chi)
\end{pmatrix}. \label{a3}
\end{eqnarray}
The determinant of the metric tensor for the space-time (\ref{a4}) is given by 
\begin{equation}
    det\,(g_{\mu\nu})=g=-\frac{\alpha^2}{4\,\Lambda^2\,f^2(\chi)\,h^6(\chi)}\,\sin^2 r\,. \label{a4}
\end{equation}

Expressing the wave equation (\ref{b1}) in the background of modified BML space-time (\ref{a2}) and using (\ref{a3})--(\ref{a4}), we obtain the following second-order differential equation:
\begin{eqnarray}
    \Bigg[-f^2(\chi)\,{\sf \frac{d^2}{dt^2}}+2\,\Lambda\,h^2(\chi)\,\Bigg\{{\sf \frac{d^2}{dr^2}}+\frac{1}{\tan r}\,{\sf \frac{d}{dr}}+\frac{1}{\alpha^2\,\sin^2 r}\,{\sf \frac{d^2}{d\phi^2}}\Bigg\}+h^2(\chi)\,{\sf \frac{d^2}{dz^2}}-M^2\Bigg]\,\Psi (t, r, \phi, z)=0\,.\label{b2}
\end{eqnarray}

In quantum mechanical system, the total wave function is always expressible in terms of different variables. Moreover, the above differential equation (\ref{b2}) is independent of time $t$, the angular coordinate $\phi$, and the translation coordinate $z$. Therefore, we choose the following wave function $\Psi (t, r, \phi, z)$ ansatz in terms of different variables as follows:
\begin{equation}
    \Psi (t, r, \phi, z)=\exp[{\sf i}\,(-E\,t+m\,\phi+k\,z)]\,\psi(r), \label{b3}
\end{equation}
where $E$ is the particle's energy, $m=0,\pm\,1,\pm\,2,....$ are the eigenvalues of the angular quantum number, and $k \geq 0$ is an arbitrary constant.

Substituting the total wave function (\ref{b3}) into the differential equation (\ref{b2}) and after separating the variables, we obtain the following differential equations for $\psi(x)$ given by
\begin{equation}
    \psi''(r)+\frac{1}{\tan r}\,\psi'(r)+\Bigg[\frac{\left(f^2(\chi)\,E^2-M^2\right)}{2\,\Lambda\,h^2(\chi)}-\frac{k^2}{2\,\Lambda}-\frac{\iota^2}{\sin^2 r}\Bigg]\,\psi(r)=0,\quad \iota=\frac{|m|}{\alpha},\label{b4}
\end{equation}
where prime denotes ordinary derivative w. r. t. $r$.

In this analysis, we are mainly interested on approximate solution of the above differential equation. However, one can try to obtain exact solution to this equation. Therefore, we write the above differential for small values of the radial distance $r$. Taking an approximation up to the first order, the radial wave equation (\ref{b4}) reduces to the following form:
\begin{equation}
    \psi''(r)+\frac{1}{r}\,\psi'(r)+\Bigg(\lambda^2-\frac{\iota^2}{r^2}\Bigg)\,\psi(r)=0,\label{c1}
\end{equation}
where we set
\begin{equation}
    \lambda^2=\frac{\left(f^2(\chi)\,E^2-M^2\right)}{2\,\Lambda\,h^2(\chi)}-\frac{k^2}{2\,\Lambda}\,\label{c22}
\end{equation}

Equation (\ref{c1}) is the Bessel second-order differential equation form whose solutions are well-known. In our case, this solution is given by $\psi(r)=c_1\,J_{\iota} (\lambda\,r)+c_2\,Y_{\iota} (\lambda\,r)$ \cite{MA,GBA}, where $J_{\iota}$ and $Y_{\iota}$, respectively are the first and second kind of the Bessel function. However, we know that the Bessel function of the second is undefined and the first kind is finite at the origin $r=0$. The requirement of wave function $\psi (r \to 0)=0$ leads to the coefficient $c_2=0$. Thus, the regular solution of the Bessel equation at the origin is given by
\begin{equation}
    \psi (r)=c_1\,J_{\iota} (\lambda\,r),\label{c2}
\end{equation}
where $c_1$ is an arbitrary constant.

The asymptotic form of the Bessel function of the first kind is given by \cite{MA,GBA}
\begin{equation}
    J_{\iota} (\lambda\,r) \propto \cos \Big(\lambda\,r-\frac{\iota\,\pi}{2}-\frac{\pi}{4}\Big).\label{c3}
\end{equation}

We aim to confine the motion of scalar particles within a region characterized by a hard-wall confining potential. This confinement is particularly significant as it provides an excellent approximation when investigating the quantum properties of systems such as gas molecules and other particles that are inherently constrained within a defined spatial domain. The hard-wall confinement is defined by a condition specifying that at a certain axial distance, $r=r_0$, the radial wave function $\psi$ becomes zero, i. e., $\psi (r=r_0)=0$. This condition is commonly referred to as the Dirichlet condition in the literature. The study of the hard-wall potential has proven valuable in various contexts, including its examination in the presence of rotational effects on the scalar field \cite{k29}, studies involving the Klein-Gordon equation under the influence of topological defect \cite{RLLV2}, examinations of a Dirac neutral particle analogous to a quantum dot \cite{RLLV6}, studies on the harmonic oscillator within an elastic medium featuring a spiral dislocation \cite{RLLV7}, and investigations into the behavior of Dirac and Klein-Gordon oscillators in the presence of a global monopole \cite{RLLV5}. This exploration of the hard-wall potential in diverse scenarios enriches our understanding of its impact on quantum systems, providing insights into the behavior of scalar particles subject to this form of confinement. Therefore, at $r=r_0$, we have $\psi(r=r_0)=0$ and using the relation (\ref{c3}) into the Eq. (\ref{c2}), we obtain the following relation:
\begin{equation}
    f^2(\chi)\,E^2=M^2+h^2(\chi)\,\Bigg[k^2+\Lambda\,\Big(2\,n+\frac{|m|}{\alpha}+\frac{3}{2}\Big)^2\,\frac{\pi^2}{2\,r^2_{0}}\Bigg],\label{c4}
\end{equation}
where $n=0, 1,2,3,....$.

By choosing different pair of rainbow function and substituting into the eigenvalue equation (\ref{c4}), one can find the approximate relativistic energy levels of scalar particles by solving the Klein-Gordon wave equation in the background of cosmological Bonnor-Melvin-type space-time with a topological defect. In this analysis, we particularly interest in two pairs of rainbow functions which are as follows and obtain the energy spectra of scalar particles using the eigenvalue relation (\ref{c4}). 

\begin{center}
\begin{figure}
\begin{centering}
\subfloat[$\alpha=0.5=\Lambda, m=1$]{\centering{}\includegraphics[scale=0.45]{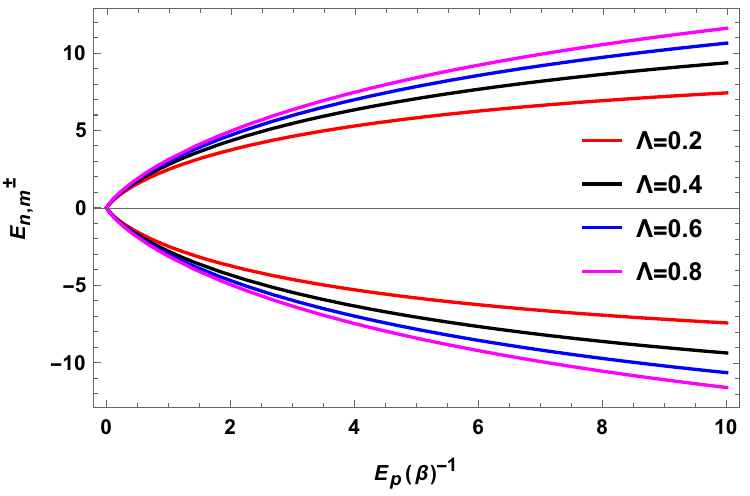}}\quad\quad
\subfloat[$\alpha=0.5,m=1,n=0$]{\centering{}\includegraphics[scale=0.45]{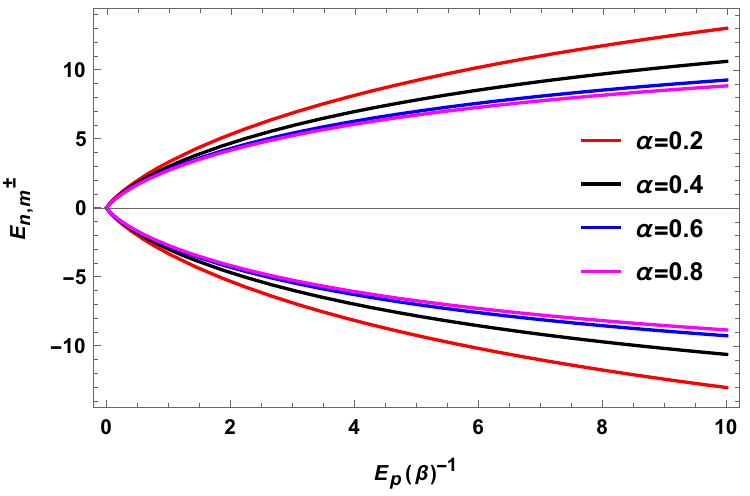}}
\par\end{centering}
\begin{centering}
\subfloat[$\Lambda=0.5, m=1, n=0$]{\centering{}\includegraphics[scale=0.45]{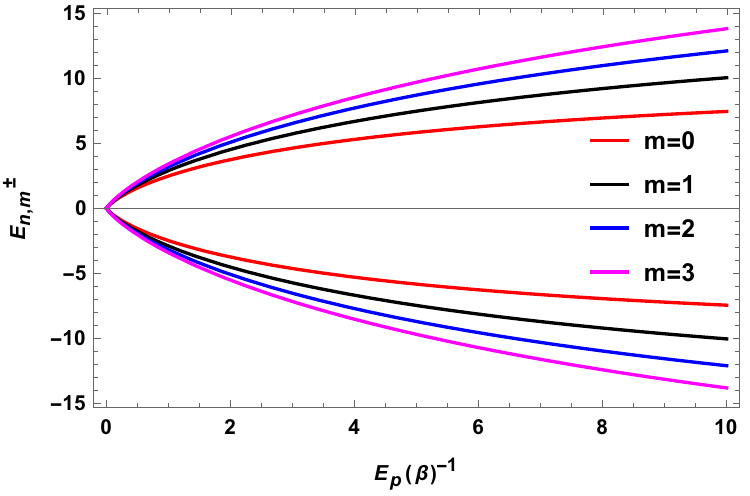}}\quad\quad
\subfloat[$\alpha=0.5=\Lambda,n=1$]{\centering{}\includegraphics[scale=0.45]{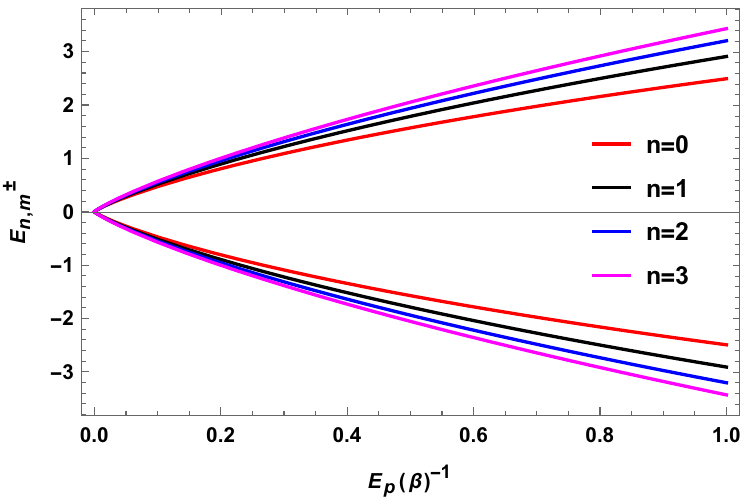}}
\par\end{centering}
\centering{}\caption{The energy spectrum $E_{n,m}^{\pm}$ for equation (\ref{c7} and \ref{c77}), where the parameters are set as $k=M=1,r=0.5.$}
\end{figure}
\par\end{center}

{\bf Case A}: Rainbow functions $f(\chi)=\frac{({\sf e}^{\beta\,\chi}-1)}{\beta\,\chi}$, \quad $h(\chi)=1$.
\vspace{0.2cm}

Here, we obtain the eigenvalue solution of the above discussed quantum system using the following pair of rainbow function given by \cite{k14}
\begin{equation}
    f(\chi)=\frac{({\sf e}^{\beta\,\chi}-1)}{\beta\,\chi},\quad h(\chi)=1, \quad \chi=\frac{|E|}{E_p}\,.\label{c5}
\end{equation}
Thereby, substituting this rainbow function into the relation (\ref{c4}), we obtain
\begin{equation}
    \frac{E^2_{p}\,\Big({\sf e}^{\frac{\beta\,|E|}{E_p}}-1\Big)^2}{\beta^2}=M^2+k^2+\Lambda\,\Big(2\,n+\frac{3}{2}+\frac{|m|}{\alpha}\Big)^2\,\frac{\pi^2}{2\,r^2_{0}}.\label{c6}
\end{equation}

For $|E|=E$, simplification of the above relation (\ref{c6}) results the following expression of the energy eigenvalue of scalar particles given by
\begin{equation}
    E^{+}_{n,m}=\frac{E_p}{\beta}\,\mbox{ln}\Bigg[1 + \frac{\beta}{E_p}\sqrt{M^2+k^2+\Lambda\,\Big(2\,n+\frac{|m|}{\alpha}+\frac{3}{2}\Big)^2\,\frac{\pi^2}{2\,r^2_{0}}}\Bigg].\label{c7}
\end{equation}
Similarly, for $|E|=-E$, simplification of the above relation (\ref{c6}) results the following expression of the energy eigenvalue of anti-particles given by
\begin{equation}
    E^{-}_{n,m}=-\frac{E_p}{\beta}\,\mbox{ln}\Bigg[1 + \frac{\beta}{E_p}\sqrt{M^2+k^2+\Lambda\,\Big(2\,n+\frac{|m|}{\alpha}+\frac{3}{2}\Big)^2\,\frac{\pi^2}{2\,r^2_{0}}}\Bigg].\label{c77}
\end{equation}

Equations (\ref{c7}), (\ref{c77}) is the approximate relativistic energy profile of charge-free scalar particles and its anti-particles in the background of Bonnor-Melvin-Lambda space-time in the presence of rainbow gravity's defined by the pair of rainbow function (\ref{c5}). We see that the relativistic energy spectrum is influenced by the topology of the geometry characterized by the parameter $\alpha$ and the cosmological constant $\Lambda$. Furthermore, the rainbow parameter $ \beta <1$ also modified the energy profile and shifted the results more.

We have represented the energy spectrum $E^{\pm}_{n, m}$ of scalar particles, as defined in Equations (\ref{c7})--(\ref{c77}), in Figure 1, while systematically varying different parameters, such as the cosmological constant $\Lambda$, the topology parameter $\alpha$, the radial quantum number $n$, and the angular quantum number $m$. These graphical illustrations reveal a consistent trend: the energy level of scalar particles in the BML-space-time background, influenced by rainbow gravity, generally increases with increasing values of these parameters, as observed in Figures 1 (a), 1 (c), and 1 (d) whereas an exceptional in Figure 1(b), where it decreases with rising values of the topological parameter $\alpha$.

\vspace{0.2cm}
{\bf Case B}: Rainbow functions $f(\chi)=1$, $h(\chi)=1+\beta\,\frac{\chi}{2}$.
\vspace{0.2cm}

In this case, we choose the following pair of rainbow function given by \cite{AFG}
\begin{equation}
    f(\chi)=1,\quad h(\chi)=1+\beta\,\frac{\chi}{2}.\label{c8}
\end{equation} 
Thereby, substituting this rainbow function into the relation (\ref{c4}), we obtain
\begin{equation}
    E^2-M^2=\Delta\,\Big(1+\frac{\beta}{2\,E_p}\,|E|\Big)^2,\label{c9}
\end{equation}
where we have set 
\begin{equation}
    \Delta=k^2+\Lambda\,\Big(2\,n+\frac{|m|}{\alpha}+\frac{3}{2}\Big)^2\,\frac{\pi^2}{2\,r^2_{0}}.\label{c10}
\end{equation}

For $|E|=E$, simplification of the above relation (\ref{c9}) results the following energy expression of scalar particles given by
\begin{equation}
    E^{+}_{n,m}=\frac{1}{\Big(1-\frac{\Delta\,\beta^2}{4\,E^2_{p}}\Big)}\,\Bigg[\frac{\Delta\,\beta}{2\,E_p} + \sqrt{M^2\,\Bigg(1-\frac{\Delta\,\beta^2}{4\,E^2_{p}}\Bigg)+\Delta} \Bigg].\label{c11}
\end{equation}
Similarly For $|E|=-E$, we obtain the energy eigenvalue of anti-particles given by
\begin{equation}
    E^{-}_{n,m}=-\frac{1}{\Big(1-\frac{\Delta\,\beta^2}{4\,E^2_{p}}\Big)}\,\Bigg[\frac{\Delta\,\beta}{2\,E_p} + \sqrt{M^2\,\Bigg(1-\frac{\Delta\,\beta^2}{4\,E^2_{p}}\Bigg)+\Delta} \Bigg].\label{c111}
\end{equation}

\begin{center}
\begin{figure}[ht]
\begin{centering}
\subfloat[$\alpha=0.5=\Lambda, m=1.$]{\centering{}\includegraphics[width=2.6in,height=2.0in]{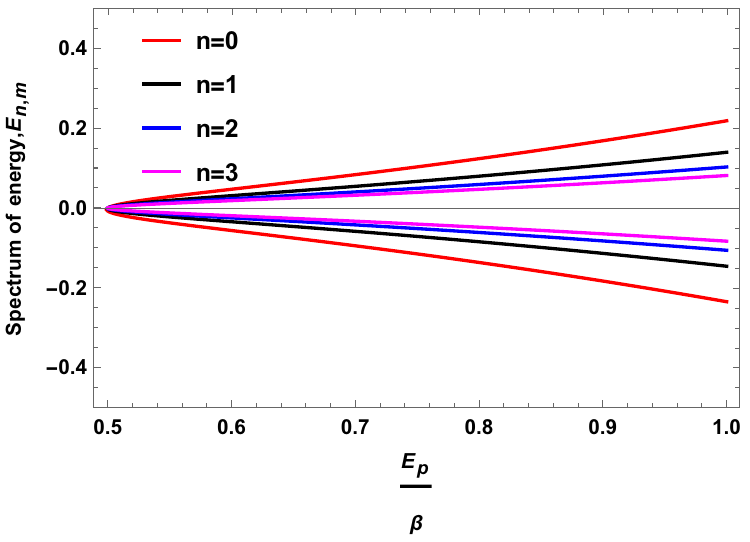}}\quad\quad
\subfloat[$\Lambda=0.5, m=1, n=0.$]{\centering{}\includegraphics[width=2.6in,height=2.0in]{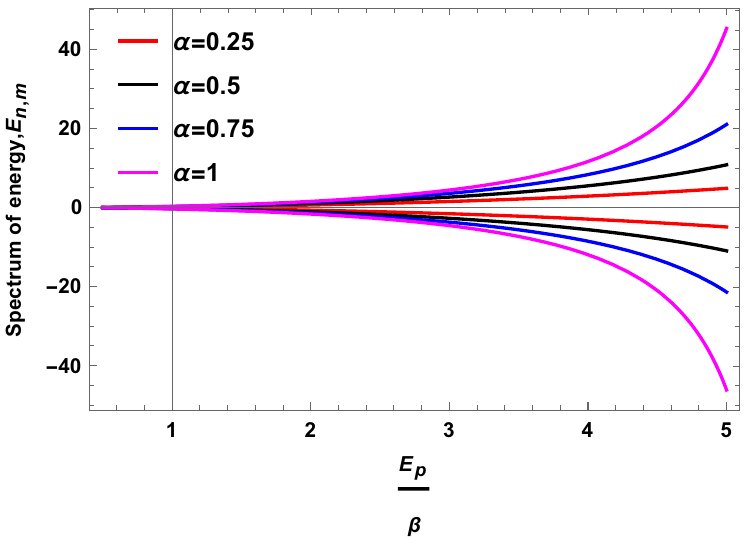}}
\par\end{centering}
\begin{centering}
\subfloat[$\alpha=0.5=\Lambda, n=0.$]{\centering{}\includegraphics[width=2.6in,height=2.0in]{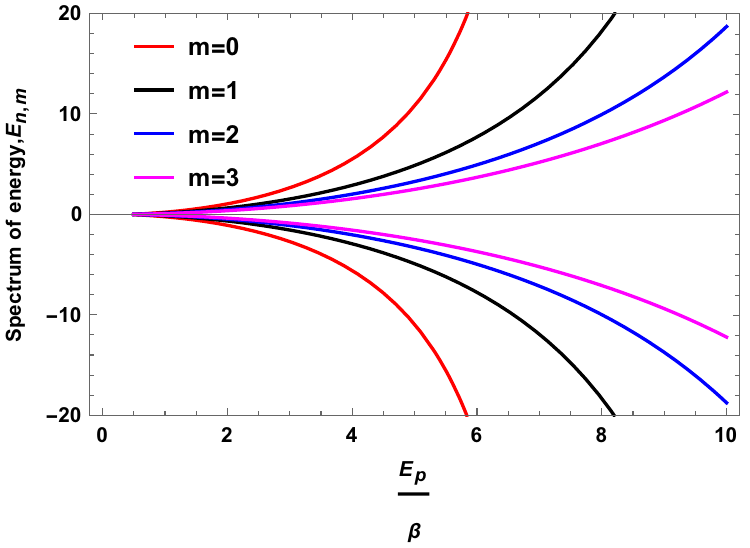}}\quad\quad
\subfloat[$\alpha=0.5,m=1, n=0.$]{\centering{}\includegraphics[width=2.6in,height=2.0in]{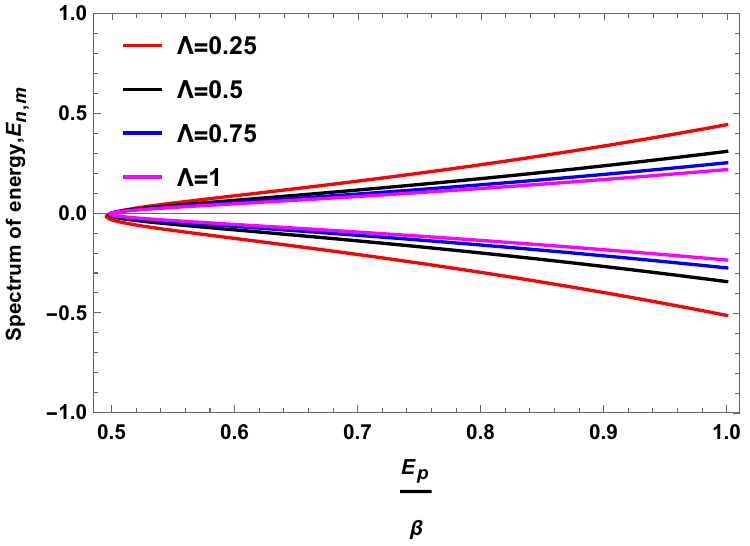}}
\par\end{centering}
\centering{}\caption{The energy spectrum $E^{\pm}_{m,n}$ for equations (\ref{c11})--(\ref{c111}). Here $k=M=1$, and $r=0.5.$}
\end{figure}
\par\end{center}

Equations (\ref{c11}), (\ref{c111}) is the approximate relativistic energy profile of charge-free scalar particles and its anti-particles in the background of Bonnor-Melvin-type cosmological space-time in the presence of rainbow gravity's defined by the pair of rainbow function (\ref{c8}). We see that the approximate energy spectrum is influenced by the topology of the geometry characterized by the parameter $\alpha$, the cosmological constant $\Lambda$, and the rainbow parameter $ \beta <1$. The presence of topological parameter breaks the degeneracy of the spectra of scalar particles and its anti-particles. 

We visually represented the energy spectrum $E^{\pm}_{n, m}$ of scalar particles, as defined in Equations (\ref{c11})--(\ref{c111}), in Figure 2, while systematically varying different parameters. These include the cosmological constant $\Lambda$, the topology parameter $\alpha$, the radial quantum number $n$, and the angular quantum number $m$. These graphical illustrations reveal a consistent trend: the energy level of scalar particles in the BML-space-time background, influenced by rainbow gravity, generally decreases with increasing values of these parameters, as observed in Figures 2 (a), 2 (c), and 2 (d). However, an intriguing exception is highlighted in Figure 2(b), where the energy level exhibits an increase with rising values of the corresponding parameter $\alpha$.

\section{Quantum Oscillator Fields: The Klein-Gordon Oscillator }

In this section, we study quantum oscillator field via the Klein-Gordon oscillator in the background of BML space-time under the influence of rainbow gravity's. This oscillator field is studied by replacing the momentum operator into the Klein-Gordon equation via $\partial_{\mu} \to (\partial_{\mu}+M\,\omega\,X_{\mu})$, where the four-vector $X_{\mu}=(0, r, 0, 0)$ and $\omega$ is the oscillator frequency. The relativistic quantum oscillator fields in curved space-times background have been investigated by numerous authors (see, Refs. \cite{RLLV5, kk28, kk29, kk30, kk31, kk32, kk33, kk35, kk35-1, kk35-2, kk35-3} and related references there in).

Therefore, the relativistic wave equation describing the quantum oscillator fields is given by
\begin{eqnarray}
    \Big[\frac{1}{\sqrt{-g}}\,(\partial_{\mu}+M\,\omega\,X_{\mu})\,(\sqrt{-g}\,g^{\mu\nu})\,(\partial_{\nu}-M\,\omega\,X_{\nu})\Big]\,\Psi=M^2\,\Psi,  \label{e1}
\end{eqnarray}
where $M$ is the rest mass of the particles, $\omega$ is the oscillator frequency, $X_{\mu}=(0, r, 0, 0)$ is the four-vector. 

Expressing the wave equation (\ref{e1}) in the magnetic universe background (\ref{a2}), we obtain
\begin{eqnarray}
    &&\Bigg[-f^2(\chi)\,{\sf \frac{d^2}{dt^2}}+2\,\Lambda\,h^2(\chi)\,\Bigg\{{\sf \frac{d^2}{dr^2}}+\frac{1}{\tan r}\,{\sf \frac{d}{dr}}-M\,\omega-\frac{M\,\omega\,r}{\tan r}-M^2\,\omega^2\,r^2+\frac{1}{\alpha^2\,\sin^2 r}\,{\sf \frac{d^2}{d\phi^2}}\Bigg\}\nonumber\\
    &&+h^2(\chi)\,{\sf \frac{d^2}{dz^2}}-M^2\Bigg]\,\Psi=0\,.\label{e2}
\end{eqnarray}

Substituting the wave function ansatz (\ref{b3}) into the above differential equation (\ref{e2}) results the following second-order differential equation form:
\begin{equation}
    \psi''+\frac{1}{\tan r}\,\psi'+\Bigg[\frac{\left(f^2(\chi)\,E^2-M^2\right)}{2\,\Lambda\,h^2(\chi)}-M\,\omega-\frac{M\,\omega\,r}{\tan r}-M^2\,\omega^2\,r^2-\frac{k^2}{2\,\Lambda}-\frac{\iota^2}{\sin^2 r}\Bigg]\,\psi=0,\label{e3}
\end{equation}
where $\iota=\frac{|m|}{\alpha}$ and prime denotes ordinary derivative w. r. t. $r$.

Following the technique adopted in the preceding section, we can write this radial wave equation (\ref{e3}) to the following form:
\begin{equation}
    \psi''+\frac{1}{r}\,\psi'+\Bigg[\eta-M^2\,\omega^2\,r^2-\frac{\iota^2}{r^2}\Bigg]\,\psi=0,\label{e4}
\end{equation}
where we have set
\begin{equation}
    \eta=\frac{\left(f^2(\chi)\,E^2-M^2\right)}{2\,\Lambda\,h^2(\chi)}-\frac{k^2}{2\,\Lambda}-2\,M\,\omega.\label{e5}
\end{equation}

We change the dependent variable via the transformation $\psi(r)=\frac{1}{\sqrt{r}}\,R(r)$, equation (\ref{e4}) can be written as the compact Liouville’s normal form
\begin{equation}
    \Bigg({\sf \frac{d^2}{dr^2}}-\sum^{2}_{i=-2}\,C_{i}\,r^{i}\Bigg)\,R(r)=0,\label{e6} 
\end{equation}
where we have set different coefficients 
\begin{eqnarray}
    C_{-2}=\iota^2-\frac{1}{4},\quad C_{-1}=0,\quad C_{0}=-\eta,\quad C_1=0,\quad C_2=M^2\,\omega^2\,.\label{e7}
\end{eqnarray}

To solve the differential equation (\ref{e6}), one can write the second term of this equation, as summation of three terms, namely: linear plus oscillator term, Coulomb plus constant term and quadratic inverse term as follows \cite{JK}:
\begin{equation}
    \sum^{2}_{i=-2}\,C_{i}\,r^{i}=V_0+V_c+V_{\infty},\label{e8}
\end{equation}
where 
\begin{equation}
    V_0=\frac{C_{-2}}{r^2},\quad V_c=C_0,\quad V_{\infty}=C_2\,r^2.\label{e9}
\end{equation}

By using equation (\ref{e8}), we can expressed the function $R(r)$ as $R(r)=A_0 (r)\,A(r)\,A_{\infty} (r)$, where $A(r)$ is an unknown function, and the functions $A_0 (r)$, $A_{\infty} (r)$ are the asymptotic factors that can be deduced from the potential functions $V_0 (r)$ and $V_{\infty} (r)$, respectively. We obtain the asymptotic factors as follows:
\begin{eqnarray}
    &&A_{0} (r)=r^{\frac{1}{2}\Big(1+\sqrt{1+4\,C_{-2}}\Big)}=r^{\iota+1/2},\nonumber\\
    &&A_{\infty} (r)=e^{-\frac{1}{2}\,\sqrt{C_2}\,\Big(r+\frac{C_1}{2\,C_2}\Big)\,r}=e^{-\frac{1}{2}\,M\,\omega\,r^2}.\label{e10}
\end{eqnarray}

Therefore, one can write the radial wave function in the following form
\begin{equation}
    R(r)=r^{\iota+1/2}\,e^{-\frac{1}{2}\,M\,\omega\,r^2}\,A(r).\label{e11}
\end{equation}
Substituting this radial function $R(r)$ using (\ref{e11}) into the differential equation (\ref{e6}), we obtain the following differential equation form:
\begin{equation}
    \Bigg({\sf \frac{d^2}{dr^2}}+\mathcal{P} (r)+\mathcal{Q} (r)\Bigg)\,A(r)=0.\label{e12}
\end{equation}
Here we have defined
\begin{eqnarray}
    \mathcal{P} (r)=\sum^{1}_{i=-1}\,p_{i}\,r^i,\quad \mathcal{Q} (r)=\sum^{0}_{i=-1}\,q_{i}\,r^i,\label{e13}
\end{eqnarray}
where the coefficients $p_{-1}, p_0, p_1, q_{-1}, q_0$  are independent of the radial coordinate $r$ and are dependent only to the potential parameters $C_{i}$, where $(-2 \leq i \leq 2)$ as follows:
\begin{eqnarray}
    &&p_{-1}=1+\sqrt{1+4\,C_{-2}}=1+2\,\iota\,\nonumber\\
    &&p_{0}=-\frac{C_1}{\sqrt{C_2}}=0,\nonumber\\
    &&p_{1}=-4\,\sqrt{C_2}=-4\,M\,\omega\,\nonumber\\
    &&q_{-1}=-\frac{C_1}{2\,\sqrt{C_2}}\,(1+\sqrt{1+4\,C_{-2}})=0,\nonumber\\
    &&q_{0}=-\sqrt{C_2}\,(2+\sqrt{1+4\,C_{-2}})=-2\,M\,\omega\,(1+\iota)\,.\label{e14}    
\end{eqnarray}

To proceed further, we express the unknown function $A(r)$ in terms of a power series expansion around the origin given by \cite{GBA} 
\begin{equation}
    A(r)=\sum^{\infty}_{i=0}\,d_i\,r^i,\label{e15}
\end{equation}
where the coefficients $d_0, d_1, d_2,.....$ depend on the parameter $C_{i}$. 

By substituting this power series (\ref{e15}) into the equation (\ref{e12}), one will find a three-term recurrence relation \footnote{This method has been discussed in details in Refs. \cite{ME,ME2}. We omitted this for simplicity and only used the energy condition in our work.} discussed in details in Refs. \cite{ME,ME2}. This power series function (\ref{e15}) becomes a finite degree polynomial by imposing the following condition (see Refs. \cite{ME,ME2} for the energy quantization condition) given by
\begin{equation}
    n\,p_1-C_0+q_0=0\quad (n=0,1,2,3,...).\label{e16}
\end{equation}
Simplification of the above condition using relations (\ref{e7}) and (\ref{e14}) gives us the following energy eigenvalue relation given by
\begin{equation}
    f^2(\chi)\,E^2=M^2+h^2(\chi)\,\Bigg[k^2+8\,M\,\omega\,\Lambda\,\Big(n+\frac{|m|}{2\,\alpha}+1\Big)\Bigg]\,.\label{e17}
\end{equation}

Here also, we use two pair of rainbow function stated in the previous section and obtain the energy eigenvalue expression of the oscillator fields.

\vspace{0.2cm}
{\bf Case A}: Rainbow functions $f(\chi)=\frac{({\sf e}^{\beta\,\chi}-1)}{\beta\,\chi}$, $h(\chi)=1$, and $\chi=\frac{|E|}{E_p}$
\vspace{0.2cm}

Substituting the pair of rainbow function $f(\chi)=\frac{({\sf e}^{\beta\,\chi}-1)}{\beta\,\chi}$ and $h(\chi)=1$ \cite{k14} into the relation (\ref{e17}), we obtain the following quadratic equation for $E$ given by
\begin{equation}
    \frac{E^2_{p}\,\Big({\sf e}^{\frac{\beta\,|E|}{E_p}}-1\Big)^2}{\beta^2}=M^2+k^2+8\,M\,\omega\,\Lambda\,\Big(n+\frac{|m|}{2\,\alpha}+1\Big)\,.\label{e18}
\end{equation}

For $|E|=E$, simplification of the above equation (\ref{e18}) results the following energy expression of scalar particles given by
\begin{equation}
    E^{+}_{n,m}=\frac{E_p}{\beta}\,\mbox{ln}\,\Bigg[1 + \frac{\beta}{E_p} \sqrt{M^2+k^2+8\,M\,\omega\,\Lambda\,\Big(n+\frac{|m|}{2\,\alpha}+1\Big)} \Bigg]\,.\label{e19}
\end{equation}

Similarly for $|E|=-E$, from the above equation (\ref{e18}), we obtain the following energy expression of anti-particles given by
\begin{equation}
    E^{-}_{n,m}=-\frac{E_p}{\beta}\,\mbox{ln}\,\Bigg[1 + \frac{\beta}{E_p} \sqrt{M^2+k^2+8\,M\,\omega\,\Lambda\,\Big(n+\frac{|m|}{2\,\alpha}+1\Big)} \Bigg]\,.\label{e199}
\end{equation}

\begin{center}
\begin{figure}
\begin{centering}
\subfloat[$\alpha=\omega=\Lambda=0.5, m=1$]{\centering{}\includegraphics[scale=0.45]{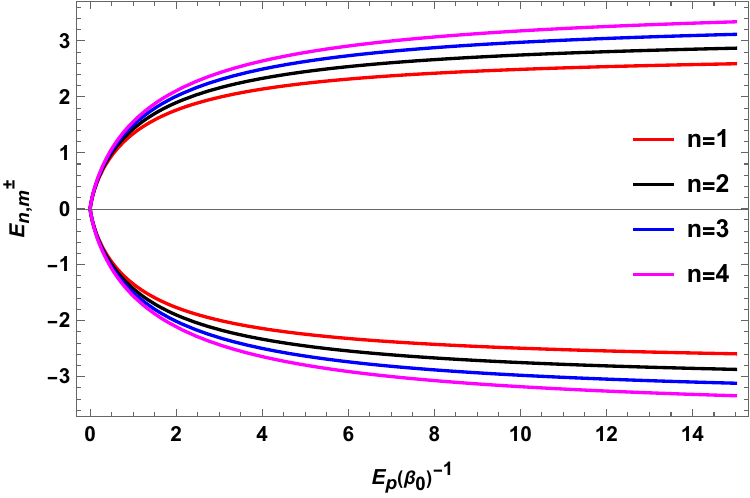}}\quad\quad\quad
\subfloat[$\alpha=\Lambda=0.5, n=1=m$]{\centering{}\includegraphics[scale=0.45]{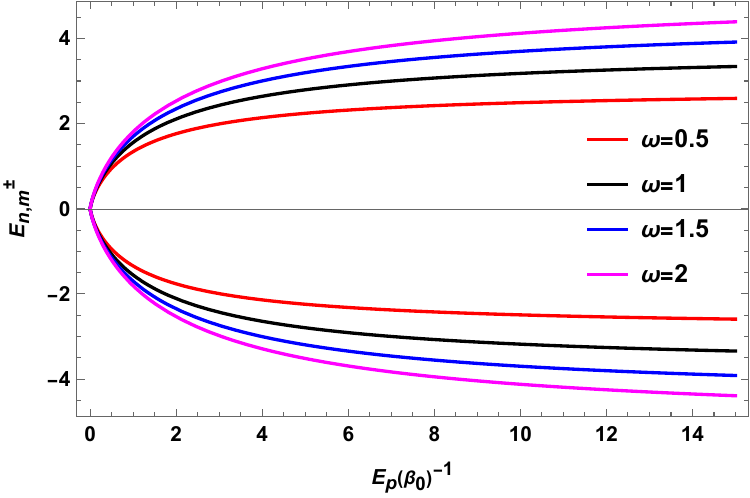}}
\par\end{centering}
\begin{centering}
\subfloat[$\alpha=\omega=0.5, n=1=m$]{\centering{}\includegraphics[scale=0.45]{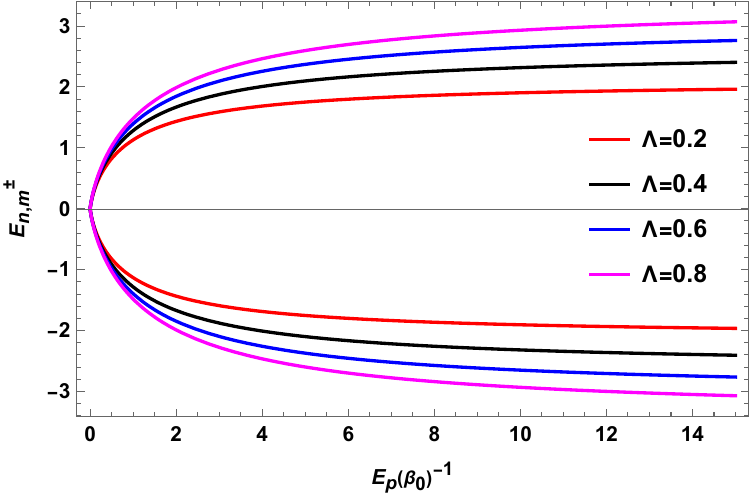}}\quad\quad\quad
\subfloat[$\omega=\Lambda=0.5, n=1=m$]{\centering{}\includegraphics[scale=0.45]{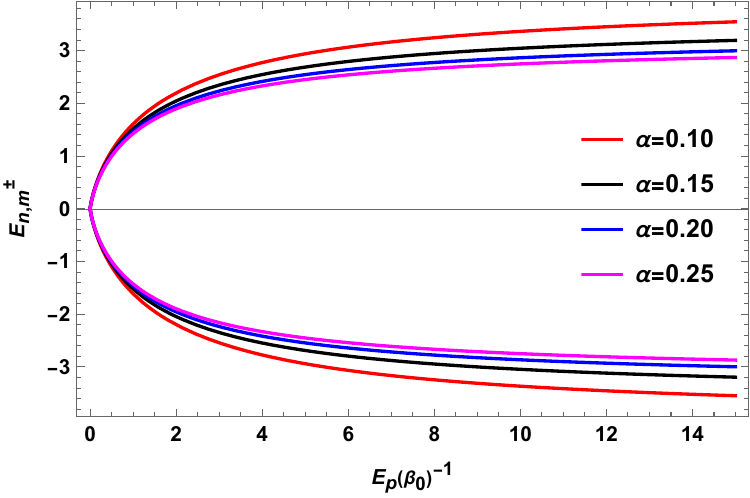}}
\par\end{centering}
\begin{centering}
\subfloat[$\alpha=\omega=\Lambda=0.5, n=1$]{\centering{}\includegraphics[scale=0.45]{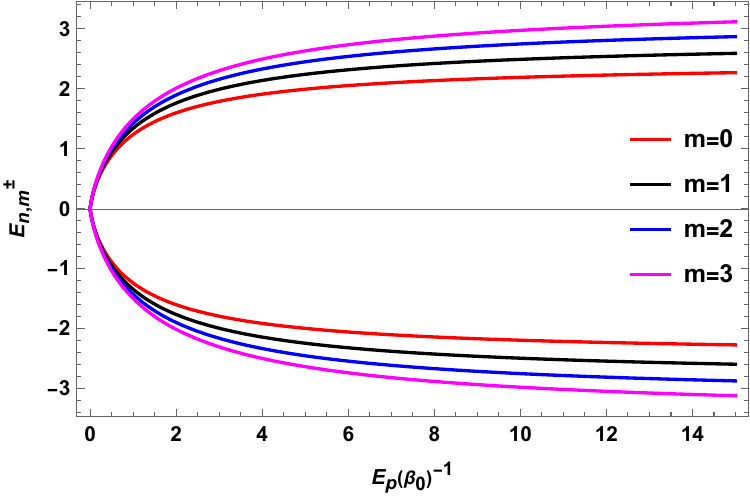}}
\par\end{centering}
\centering{}\caption{The energy spectrum $E_{m,n}^{\pm}$ for the relation given in equation (\ref{e19} and \ref{e199}). Here, $k=M=1$. }
\end{figure}
\par\end{center}

Equations (\ref{e19}), (\ref{e199}) is the approximate relativistic energy profile of oscillator field and its anti-particles in the background of Bonnor-Melvin-Lambda space-time in the presence of rainbow gravity's defined by the pair of rainbow function (\ref{c5}). We see that the approximate energy spectrum is influenced by the topology of the geometry characterized by the parameter $\alpha$, the cosmological constant $\Lambda$, and changes with change in the oscillator frequency $\omega$. Furthermore, the rainbow parameter $ \beta <1$ also modified the energy profiles and shifted the results more. One can see that the presence of topological parameter breaks the degeneracy of the spectra of energy of oscillator field. 

We have generated Figure 3 to illustrate the energy spectra, $E^{\pm}_{n, m}$, as a function of $E_p/\beta$ for various parameter dependencies. Specifically, Figure 3(a) demonstrates the variation with respect to the quantum number $n$, while Figure 3(b) showcases the impact of the oscillator frequency $\omega$. In Figure 3(c), we show how the energy spectra change in response to alterations in the cosmological constant $\Lambda$, and in Figure 3(d), the behavior of energy level with respect to the topological parameter $\alpha$. Additionally, Figure 3(e) investigates the influence of the angular quantum number $m$. Our findings reveal a consistent trend across most parameters: as their values increase, so does the energy level, as depicted in Figures 3(a), 3(b), 3(c), and 3(e). However, it is noteworthy that this pattern is inverted for the topological parameter $\alpha$, as shown in Figure 3(d), where the energy level decreases with increasing values of $\alpha$.

\vspace{0.2cm}
{\bf Case B}: Rainbow functions $f(\chi)=1$, $h(\chi)=1+\beta\,\frac{\chi}{2}$, and $\chi=\frac{|E|}{E_p}$
\vspace{0.2cm}

\begin{center}
\begin{figure}
\begin{centering}
\subfloat[$\alpha=\omega=\Lambda=0.5, m=1$]{\centering{}\includegraphics[scale=0.45]{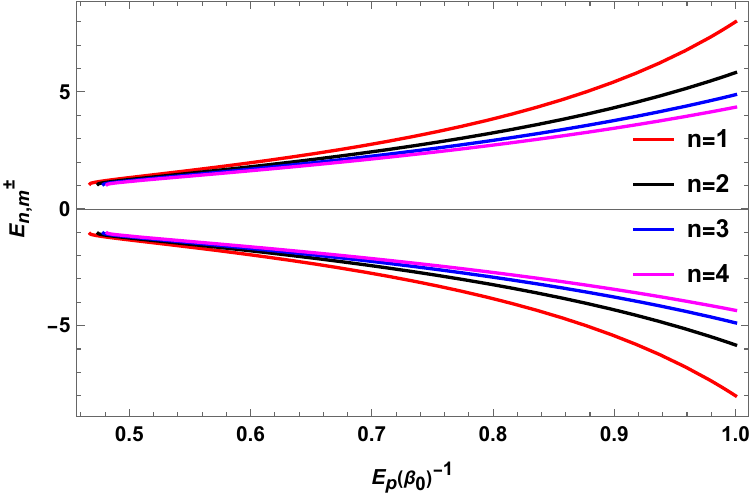}}\quad\quad\quad
\subfloat[$\alpha=\Lambda=0.5, n=1=m$]{\centering{}\includegraphics[scale=0.45]{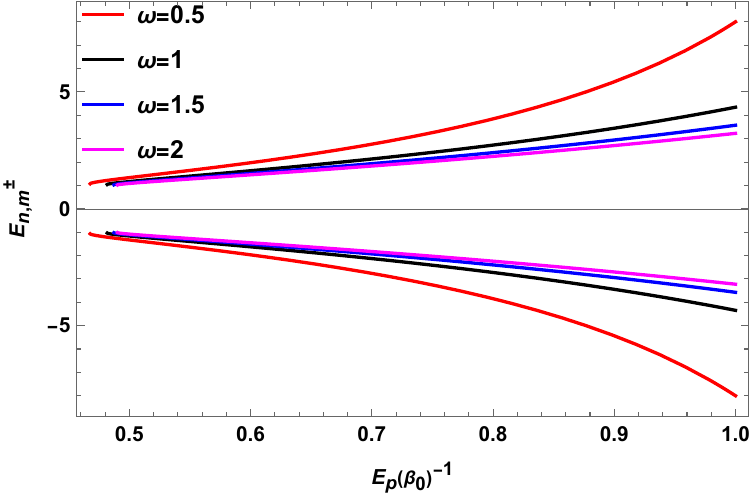}}
\par\end{centering}
\begin{centering}
\subfloat[$\alpha=\omega=0.5, n=1=m$]{\centering{}\includegraphics[scale=0.45]{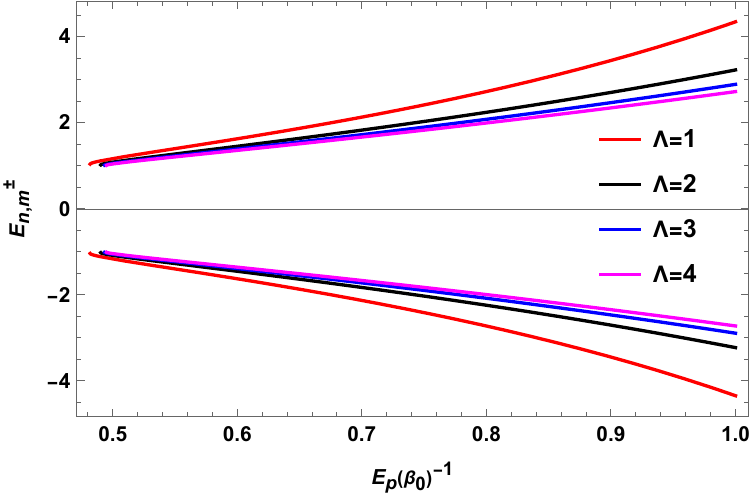}}\quad\quad\quad
\subfloat[$\omega=\Lambda=0.5, n=1=m$]{\centering{}\includegraphics[scale=0.45]{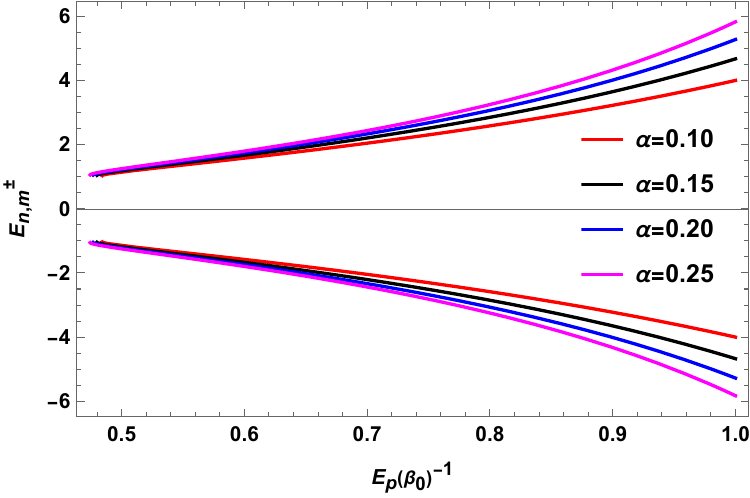}}
\par\end{centering}
\begin{centering}
\subfloat[$\alpha=\omega=\Lambda=0.5, n=1$]{\centering{}\includegraphics[scale=0.45]{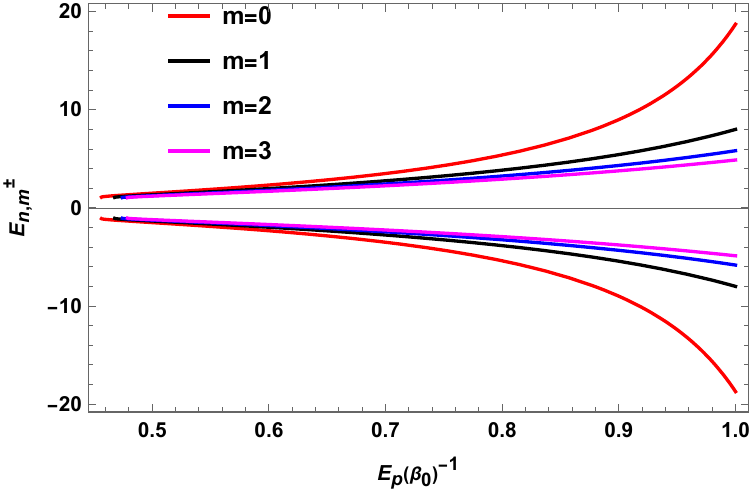}}
\par\end{centering}
\centering{}\caption{The energy spectrum $E_{m,n}^{\pm}$ for the relation given in equation (\ref{e22} and \ref{e222}). Here, $k=M=1$. }
\end{figure}
\par\end{center}

Substituting the pair of rainbow function $f(\chi)=1$, $h(\chi)=1+\beta\,\frac{\chi}{2}$ \cite{AFG} into the relation (\ref{e17}), we obtain the following quadratic equation for $E$ given by
\begin{equation}
    E^2-M^2=\Theta\,\Big(1+\frac{\beta}{2\,E_p}\,|E|\Big)^2,\label{e20}
\end{equation}
where we set
\begin{equation}
    \Theta=k^2+8\,M\,\omega\,\Lambda\,\Big(n+\frac{|m|}{2\,\alpha}+1\Big)\,.\label{e21}
\end{equation}

For $|E|=E$, simplification of the above equation (\ref{e20}) results the following energy expression of oscillator field associated with the mode $\{n, m\}$ given by
\begin{equation}
    E^{+}_{n,m}=\frac{1}{\Bigg[1-\frac{\Theta}{4}\,\Big(\frac{\beta}{E_p}\Big)^2\Bigg]}\,\Bigg[\frac{\Theta}{2}\,\frac{\beta}{E_p} + \sqrt{M^2\,\Bigg\{1-\frac{\Theta}{4}\,\Big(\frac{\beta}{E_p}\Big)^2\Bigg\}+\Theta} \Bigg].\label{e22}
\end{equation}
Similarly for $|E|=-E$, we obtain the energy eigenvalue of anti-particles associated with the mode $\{n, m\}$ given by
\begin{equation}
    E^{-}_{n,m}=-\frac{1}{\Bigg[1-\frac{\Theta}{4}\,\Big(\frac{\beta}{E_p}\Big)^2\Bigg]}\,\Bigg[\frac{\Theta}{2}\,\frac{\beta}{E_p} + \sqrt{M^2\,\Bigg\{1-\frac{\Theta}{4}\,\Big(\frac{\beta}{E_p}\Big)^2\Bigg\}+\Theta} \Bigg].\label{e222}
\end{equation}

Equations (\ref{e22}), (\ref{e222}) is the approximate relativistic energy profile of oscillator field in the background of Bonnor-Melvin-Lambda space-time in the presence of rainbow gravity's defined by the pair of rainbow function (\ref{c8}). We see that energy spectrum is influenced by the topology of the geometry characterized by the parameter $\alpha$, the cosmological constant $\Lambda$,  and the oscillator frequency $\omega$. Furthermore, the rainbow parameter $ \beta <1$ also modified the energy profiles and shifted the results.

We've produced Figure 4 to depict the energy spectra, $E^{\pm}_{n, m}$, in relation to $E_p/\beta$ across various parameter variations. In Figure 4(a), we analyze the dependency on the quantum number $n$, while Figure 4(b) explores the influence of the oscillator frequency $\omega$. Figure 4(c) illustrates the impact of changes in the cosmological constant $\Lambda$, and in Figure 4(d), we examine the behavior concerning the topological parameter $\alpha$. Additionally, Figure 4(e) investigates the effect of the angular quantum number $m$. Our analysis reveals a consistent trend across most parameters: as their values increase, the energy level tends to decrease, as demonstrated in Figures 4(a), 4(b), 4(c), and 4(e). However, it's noteworthy that this trend is reversed for the topological parameter $\alpha$, as depicted in Figure 4(d), where the energy level increases with rising values of $\alpha$.

\section{Conclusions}

Rainbow gravity, an intriguing phenomenon in physics, introduces modifications to the relativistic mass-energy relation within the framework of special relativity theory. Consequently, the exploration of the implications of rainbow gravity in quantum mechanical problems has emerged as a significant area of research interest. Numerous studies have investigated the effects of rainbow gravity on diverse quantum systems, showcasing its impact on various phenomena. These studies span a range of quantum systems, including the Dirac oscillator in cosmic string space-time \cite{aa13}, the dynamics of scalar fields in a wormhole background with cosmic strings \cite{AG}, quantum motions of scalar particles \cite{EE}, and the behavior of spin-1/2 particles in a topologically trivial Gödel-type space-time \cite{EE2}. Investigations have also extended to the motions of photons in cosmic string space-time \cite{EE3} and the generalized Duffin–Kemmer–Petiau equation with non-minimal coupling in cosmic string space-time \cite{EE4}. 

The effects of rainbow gravity have also been investigated by Ali \textit{et al.} in various astrophysical phenomena, such as black holes \cite{xxx1}, gravitational collapse \cite{xxx2}, the formation of remnants for all black objects (Kerr, Kerr–Newman-dS, charged-AdS, and higher-dimensional Kerr–AdS black holes) \cite{xxx3}, the absence of black holes in LHC \cite{xxx4}, investigation of spinning and charged black rings \cite{xxx5}, FRW cosmologies \cite{xxx6}, and the initial singularity problem \cite{xxx7}. In Ref. \cite{xxx5}, the authors calculated corrections to the temperature, entropy, and heat capacity of black rings and demonstrated that Hawking radiation changes considerably near the Planck scale in gravity's rainbow, where black rings do not evaporate completely and a remnant is left as the black rings evaporate down to the Planck scale. In our present study, we explored the quantum dynamics of scalar particles under the influence of rainbow gravity in a four-dimensional curved geometry background known as Bonnor-Melvin-Lambda space-time, featuring a cosmological constant. This investigation specifically focused on examining the behavior of quantum particles in the system under consideration. By extending the scope of research to this unique space-time background, we contributed to the growing body of knowledge surrounding the intricate interplay between rainbow gravity and quantum systems in this magnetic field background space-time.

In {\it Section 2}, we derived the radial equation of the Klein-Gordon equation, describing the dynamics of charge-free spin-0 scalar particles within the framework of BML space-time under the influence of rainbow gravity. Subsequently, we solved this radial equation by selecting two pairs of rainbow functions, resulting in approximate energy eigenvalues for scalar particles, given by Equations (\ref{c7})--(\ref{c77}) and (\ref{c11})--(\ref{c111}). The energy of scalar particles is denoted as $E^{+}_{n,m}$, whereas $E^{-}_{n,m}$ represents its antiparticles. We have observed that the values of energies for particles and antiparticles are the same and are equally spaced on either side of $E=0$, as observed in Figures 1-2. We have shown that with increasing values of a few parameters, the energy level increases, and in some cases, it decreases.

Moving on to {\it Section 3}, we explored the quantum oscillator fields through the Klein-Gordon oscillator in curved space-time under consideration, influenced by rainbow gravity. We derived the radial equation of the relativistic wave equation and, employing the same pairs of rainbow functions, obtained the approximate energy profiles of the Klein-Gordon oscillator field, as outlined in Equations (\ref{e19})--(\ref{e199}) and (\ref{e22})--(\ref{e222}). In both studies, we have demonstrated that the energy profiles of both scalar particles and oscillator fields are influenced by the topology parameter of curved geometry characterized by the parameter $\alpha$, the cosmological constant $\Lambda$, and the quantum numbers $\{n, m\}$. Additionally, in the context of quantum oscillator fields, the frequency of oscillation $\omega$ also modifies the energy spectrum of oscillator fields and shifts the results further. Another noteworthy aspect is that the energy spectrum for both scalar particles and oscillator fields depends on the topological parameter $\alpha$, which breaks the degeneracy of the spectra of energy. To demonstrate the influence of various parameters involved in the energy expressions in both studies, we visually represented these energy spectra for oscillator fields across various values of the aforementioned parameters in Figures 3--4. We have shown that with increasing values of a few parameters, the energy level increases, and in some cases, it decreases.

Our research findings shed light on the intricate relationship between the dynamics of quantum scalar particles and the magnetic field within the Bonnor-Melvin-Lambda space-time, featuring a topological parameter analogous to a cosmic string. This exploration offers valuable insights into the behavior of both scalar particles and oscillator fields when exposed to the effects of rainbow gravity. The study enhances our comprehension of how rainbow gravity impacts quantum systems within the particular geometric framework of BML space-time.

\section*{Conflict of Interest}

There is no conflict of interests in this paper.

\section*{Funding Statement}

No fund has received for this paper.

\section*{Data Availability Statement}

No data were generated or analysed during this study.

\end{document}